# Dependence of the $0.5\times(2e^2/h)$ conductance plateau on the aspect ratio of InAs quantum point contacts with in-plane side gates

P. P. Das[1,(a)], A. Jones[2], M. Cahay[2,(a)], S. Kalita[3], S. S. Mal[4], N. S. Sterin[1], T. R. Yadunath[1,5], M. Advaitha[1,5], and S. T. Herbert[6]

[1]Department of Physics, National Institute of Technology Karnataka,
Surathkal 575025, India
[2]Spintronics and Vacuum Nanoelectronics Laboratory,
University of Cincinnati, Cincinnati, Ohio  45040, USA
[3]Department of Electronics and Communication Engineering, Tezpur University, Assam 784028, India
[4]Department of Chemistry, National Institute of Technology Karnataka,
Surathkal 575025, India
[5]Center for Nano Science and Engineering (CeNSE),
Indian Institute of Science Bangalore, Bangalore 560012, India
[6]Department of Physics, Xavier University, Cincinnati, Ohio 45207, USA

**Abstract**

The observation of a $0.5\times(2e^2/h)$ conductance plateau in asymmetrically biased quantum point contacts (QPCs) with in-plane side gates (SGs) has been attributed to the onset of spin-polarized current through these structures. For InAs QPCs with the same width but longer channel length, there is roughly a fourfold increase in the range of common sweep voltage applied to the SGs over which the $0.5\times(2e^2/h)$ plateau is observed when the QPC aspect ratio (ratio of length over width of the narrow portion of the structure) is increased by a factor 3. Non-equilibrium Green's function simulations indicate that the increase in the size of the $0.5\times(2e^2/h)$ plateau is due to an increased importance, over a larger range of common sweep voltage, of the effects of electron-electron interactions in QPC devices with larger aspect ratio. The use of asymmetrically biased QPCs with in-plane SGs and large aspect ratio could therefore pave the way to build robust spin injectors and detectors for the successful implementation of spin field effect transistors.

(a) Corresponding authors: daspm@nitk.ac.in and marc.cahay@uc.edu

## 1. Introduction

In recent years, spin-orbit coupling (SOC) has been established as a possible tool for all-electrical spin control and generation of spin-polarized currents[1-4]. Many recent experimental reports have shown that an asymmetry in the potential applied between the two side gates (SGs) of a quantum point contact (QPC) can be used to create strongly spin polarized currents using either top or in-plane SGs[5-13]. Recently, Chuang et al. demonstrated the operation of an all-electric spin valve by fabricating near 100% efficient spin injector and detector using asymmetrically biased QPCs with top gates placed on top of an AlGaAs/GaAs heterostructure[14]. Furthermore, Chuang et al. have shown that control of spin precession in the semiconducting channel between the two QPCs can be achieved using a middle gate to tune the strength of the Rashba SOC in the channel while maintaining ballistic transport to minimize the effects of spin decoherence in that region. Operation of such a spin-valve at higher temperatures may be achieved by incorporating in-plane SGs between the spin injector and detector to the device architecture[15].

The effects of electron-electron interactions were hard to assess in the QPCs used by Chuang et al.[14]. In the past, we have shown that *lateral* spin-orbit coupling (LSOC), resulting from the lateral in-plane electric field of the confining potential of InAs- and GaAs-based QPCs with in-plane SGs, can be used to create a strongly spin-polarized current by purely electrical means[10-13]. This was realized by tuning the electron confinement potential of the channel using asymmetric bias voltages on the SGs. It is noteworthy that our QPC structures that exploit only LSOC are clearly different from the conventional top-gated (or split-gated) ones[16-22]. In top-gated QPCs it is the Rashba SOC, rather than LSOC, which controls the electron spin in 2DEG. LSOC arises due to in-plane bias asymmetry on the SGs whereas Rashba SOC results from an electric field in the direction of growth of an *asymmetric* quantum well. Since, a nominally

*symmetric* InAlAs/InAs quantum well structure is used in this work, the presence of Rashba SOC can be ruled out in our QPC devices and LSOC is the leading mechanism affecting spin transport through the narrow portion of the QPC.

In some of our earlier works we observed a conductance plateau at $0.5 \times (2e^2/h)$ in the conductance plots of both InAs[10,11,13] and GaAs[12] asymmetrically biased side-gated QPCs. A Non-Equilibrium Green's Function (NEGF) analysis was used to model a small QPC and three sufficient ingredients were found to generate a strong spin polarization[10,23,24]: (1) an asymmetric lateral confinement in the QPC channel, (2) a LSOC induced by the lateral confining potential of the QPC and (3) a strong electron-electron (e-e) interaction[10-13,23,24].

Some of our previous experimental and theoretical results have also found the presence of other conductance anomalies (i.e., at conductance values different from $0.5 \times (2e^2/h)$)[13]. The main reason for these occurrences was shown to be due to the influence of defects (surface roughness and impurity or dangling bond scattering from the rugged QPC walls), as can be seen in the scanning electron micrograph (SEM) of a QPC shown in Fig. 1. These anomalies are believed to be signatures of spin polarization in the QPC, triggered by the imbalance between the lateral spin-orbit coupling on opposite sidewalls due to the applied asymmetric potential, $\Delta V_G$ between the two SGs.

In this report, we show that InAs QPCs with in-plane SGs can be used as efficient and robust spin injectors and detectors by increasing the aspect ratio of the QPC channel (ratio of length to width of the narrow portion of the QPC). We demonstrate that the $0.5 \times (2e^2/h)$ anomalous conductance plateau observed in these QPCs can be achieved over a larger range of common sweep voltage on the SGs while increasing the QPC aspect ratio. The appearance of a 0.5 conductance anomaly in QPCs with top gates as the length of the narrow portion is increased was demonstrated theoretically in the past by Jaksch et al.[25] and was an incentive for this work.

## 2. Experiments

In this study, a high mobility 2DEG was formed using a modulation doped symmetric InAlAs/InAs quantum well (QW) grown by molecular beam epitaxy. In the heterostructure, a thin layer of InAs (low bandgap ~ 0.36 eV) is sandwiched between two layers of InGaAs (large bandgap ~ 0.76 eV), thus forming a quantum well (QW), as shown in Fig. 2. The QW structure was grown on a thick semi-insulating InP substrate (shown in Fig. 2). A 300 nm InAlAs buffer layer was grown on the substrate to compensate for the lattice mismatch. The 7 nm n-doped InAlAs supply layer sits on the buffer layer and supplies electrons to the QW. The active layers that form the two-dimensional electron gas (2DEG) are separated from the supply layer by a 10 nm spacer layer. The 2DEG was formed by sandwiching a thin (thickness of 3.5 nm) InAs layer between two InGaAs layers. On the top, a 2 nm InAs cap layer was deposited to prevent oxidation and other physical damage of the active layers underneath. For InAs the Fermi level is pinned far above the conduction band minimum making it easier to deposit ohmic contacts on it. Shubnikov-de Haas and quantum Hall measurements performed in dark condition at 4.2 K on a simple Hall bar structure yielded carrier concentration and electron mobility of the 2DEG equal to $2.2 \times 10^{16}/m^2$ and $3.67\ m^2/Vs$, respectively.

The wafer was cleaned in hot acetone, methanol and isopropanol (for 10 minutes each), then washed in an oxygen plasma for 40 seconds. Next, it was pre-etched in 4% HCl for 5 minutes, rinsed in deionized water, and pre-baked at 185 ºC for 5 minutes. A 50 nm thick polymethylcrylate (PMMA) electron beam resist was spin-coated and then exposed, using electron beam lithography, to write the QPC pattern. The electron dose was 65 $\mu C/cm^2$ and the voltage 10 kV. The pattern was then developed in MIBK:isopropanol (1:1) for 65 seconds. After post-baking the sample at 115 $^0$C for 5 minutes, it was wet-etched in

H$_2$O:H$_2$O$_2$:CH$_3$COOH (125:20:15) for 25 seconds to form two trenches that define the in-plane side gates (SGs). Contact pads (12 nm Ni, 20 nm Ge and 300 nm Au) were deposited on top of the InAs cap layer using thermal and electron-beam evaporation technique, followed by a rapid thermal annealing at 350 $^0$C for 120 second to form ohmic contacts with the 2DEG underneath.

The QPC constrictions were defined by cutting deep trenches through the 2DEG using electron beam lithography followed by wet etching technique. The deep trenches punching through the 2DEG leads to stronger LSOC than achievable in QPC devices with shallow trenches[26,27]. Five QPCs with varying aspect ratio were fabricated. A SEM image of one of our fabricated QPCs with two in-plane side gates (G1 and G2) is shown in Fig.1. The narrow portion of QPCs 1 through 5 had the same lithographic width W (around 270 nm) but a different length L equal to 320 nm, 380 nm, 510 nm, 820 nm and 930 nm, respectively. This corresponds to a QPC aspect ratio (L/W) ranging from 1.2, 1.4, 1.9, 3.0, to 3.4 for QPC 1 through 5, respectively. Since InAs is a material with high intrinsic SOC, it has a short spin coherence length of about a micron at 4.2 K[28]. This is the reason the maximum length of the QPC channel was kept less than one micron. The electrostatic width of the QPC channel was changed by applying bias voltages to the metallic in-plane SGs gates, depleting the channel near the sidewalls of the QPC. Battery operated DC voltage sources were used to apply constant negative voltages $V_{G1}$ and $V_{G2}$ to the two SGs (as shown in Fig. 3). An asymmetric potential $\Delta V_G = V_{G1} - V_{G2}$ between the two gates was applied to create spin polarization in the channel. In all conductance measurements the contact pads showed ohmic character and the side-gates were found to be non-leaky at 4.2 K. For all conductance measurements the QPC devices were first cooled down to liquid helium temperature (4.2 K). Figure 3 shows circuit diagram used in the conductance measurements of all the QPCs.

Since there is very little surface depletion at the InAs 2DEG/vacuum interface at 4.2K, a conducting channel through the QPC narrow constriction can be realized with a very narrow channel width. This is an added advantage for InAs side-gated QPCs. The lithographic width of the QPC channel was varied by applying bias voltages to the metallic in-plane SGs, depleting the channel near the sidewalls of the QPC. The asymmetric potential, $\Delta V_G = V_{G1} - V_{G2}$ between the two SGs was applied to create spin polarization in the channel[10-13]. The QPC conductance was then recorded as a function of a common sweep voltage, $V_G$, applied to the two SGs in addition to the potentials $V_{G1}$ and $V_{G2}$, with the current flowing in the x-direction (Fig. 3). The linear conductance $G$ (= $I/V_{ds}$) of the channel was measured for different $\Delta V_G$ as a function of $V_G$, using a four-probe lock-in technique with a drive frequency of 17 Hz and a drain-source drive voltage of $V_{ds}$ = 100 µV. The value of $\Delta V_G$ was adjusted until a robust 0.5×(2$e^2$/h) plateau was detected. As shown previously[10-13], the 0.5×(*2e²/h*) anomaly does not arise when the applied potential on the SGs was identical (i.e., when the SGs were *symmetrically* bias), a feature also supported by our earlier NEGF simulations[23,24].

## 3. Results

Figure 4 shows plots of the conductance $G$ (in units of *2e²/h*) of the five QPCs as a function of the common sweep voltage $V_G$ applied to the in-plane SGs. Figure 4 also shows that the normal conductance plateau in almost all five cases is not located exactly at conduction value *2e²/h*, a feature attributed to the influence of surface scattering in our earlier finding[13] (due to dangling bonds and surface roughness in the narrow portion of the QPC). As we showed earlier[11], the presence of surface scattering in QPCs with in-plane SGs can be illustrated by performing the conductance measurements in a magnetic field perpendicular to the device plane or the 2DEG. By increasing the strength of the external magnetic field, it has been found that the anomalous plateau evolves smoothly towards the normal conductance plateau at *2e²/h*[11]. This is due to the fact that magnetic confinement helps diminish scattering from the QPC side walls. Transport through the channel is then near-ballistic and the normal conductance plateau becomes well-defined[11]. There are also some small fluctuations in the conductance curves which are more clearly seen in the plot of the transconductance curves, (dG/dV$_G$), also shown in Fig.4. Besides surface scattering, another likely source for the small fluctuations is the influence of the donor supply layer which is located only 10 nm away from the active well. Indeed, recent modeling suggests that such a short separation may lead to substantial fluctuation in the conductance of QPCs[29].

For clarity and convenience for comparing the flatness of the plateau near the 0.5×(2e$^2$/h) conductance anomaly, some of the conductance plots were laterally shifted along the common sweep voltage ($V_G$) axis and the range of V$_G$ was kept between -7 V and -1 V for all five QPCs. In Fig. 4, the transconductance plots were calculated numerically after smoothening out the conductance data. The transconductance plot (*dG/dV*$_G$ vs V$_G$) was used to extract the range of the common sweep voltage over which the 0.5×(*2e$^2$/h*) plateau is observed using the procedure illustrated in Fig. 5. First, the location of the common sweep voltage at which the transconductance plot shows a (non-zero) minimum, $(dG/dV_G)_{min}$, near the 0.5×(*2e$^2$/h*) plateau was measured. This value is represented by a horizontal dotted red line in Fig. 5. A second dotted green horizontal line was then drawn at a value corresponding to three times $(dG/dV_G)_{min}$. The values of the common sweep voltages (V$_1$ and V$_2$) at which the second horizontal line crossed the transconductance curve were then used to calculate the flatness of the conductance curve near the 0.5×(*2e$^2$/h*) conductance plateau as follows, $\Delta V_G^{Flat} = |V_2 - V_1|$. Following this procedure, the extracted values of $\Delta V_G^{Flat}$ for the five QPCs were found to be equal to 0.3, 0.5, 0.8, 1.0, and 1.2V, for QPC 1, 2, 3, 4, and 5, respectively. QPC 5 with the largest aspect ratio has the largest value of $\Delta V_G^{Flat}$. The latter decreases gradually from QPC 4 to QPC 1. The arbitrariness of the factor 3 used to define the horizontal red and green lines shown in Fig. 5 can be alleviated by comparing the relative sizes of $\Delta V_G^{Flat}$ for the five different QPCs with QPC 1 selected as a reference. The results of this comparison are listed in Table 1. This table shows a near 4-fold increase in the value of $\Delta V_G^{Flat}$ with a near

3-fold increase of the QPC aspect ratio. This indicates that the onset of spin polarization can be achieved over larger range of the common sweep voltage in QPCs with larger aspect ratio. This assertion is supported by the NEGF simulations described next.

## 4. Discussion

The NEGF approach was used to study the appearance and shape of the conductance anomalies as a function of the aspect ratio of a QPC with in-plane SGs. The model QPC used is shown in Fig. 6, where the white region represents the QPC channel. The black areas represent the deep-etched isolation trenches that define the lithographic dimensions of the QPC constriction. The gray areas represent the sharp potential walls between the inside of the QPC and the isolation trenches. Also shown are the electrodes connected to the QPC device: source, drain and two SGs.

The QPC shown in Fig. 6 was assumed to be made from a nominally symmetric InAlAs/InAs quantum well. Spatial inversion asymmetry is therefore assumed to be negligible along the growth axis ($z$ axis) of the QW and the corresponding Rashba spin-orbit interaction is neglected. The Dresselhaus spin-orbit interaction due to the bulk inversion asymmetry in the direction of current flow was also neglected. The only spin-orbit interaction considered is the lateral spin-orbit coupling (LSOC) due to the lateral confinement of the QPC channel, provided by the isolation trenches and the bias voltages of the SGs[10-13]. The free-electron Hamiltonian of the QPC is given by

$$H_0^\sigma = H_0 I + \frac{\beta}{\hbar} \vec{\sigma} \cdot (\vec{p} \times \vec{\nabla} U), \qquad (1)$$

where $H_0 = \frac{1}{2m^*}(p_x^2 + p_y^2) + U(x, y)$, $I$ is the 2x2 identity matrix, $\vec{p}$ the momentum operators in the x-y plane, β is intrinsic SOC parameter characterizing the strength of the LSOC, $\vec{\sigma}$ is the vector of Pauli spin matrices and U(x, y) the potential energy representing the sum of conduction band energy profile within and around the QPC and effects of space-charge through a solution of the Poisson equation.

The low band gap semiconductor InAs has a large intrinsic SOC. The effective mass in the InAs channel was set equal to $m^* = 0.023 m_0$, where $m_0$ is the free electron mass. The 2DEG is assumed to be located in the (x, y) plane, x being the direction of current flow from source to drain and y the direction of transverse confinement of the channel. U(x, y) is the confinement potential, which includes the potential introduced by gate voltages and the conduction band discontinuity at the InAs/air interface.

The conductance through the QPC was calculated using a NEGF method under the assumption of ballistic transport[23,24]. We used a Hartree-Fock approximation following Lassl et al.[30] to include the effects of electron-electron (e-e) interaction in the QPC. More specifically, the e-e interaction was taken into account by considering a repulsive Coulomb contact potential, $V_{int}(x,y\,;\,x',y') = \gamma\, \delta(x-x')\, \delta(y-y')$, where $\gamma$ indicates the e-e interaction strength. As a result, an interaction self-energy, $\Sigma_{int}^{\sigma}(\vec{r}) = \gamma n_{-\sigma}(\vec{r})$, must be added to the Hamiltonian in Eq.(1)[23,24]. The interaction self-energy $\Sigma_{int}^{\sigma}(x,y)$ is different for the two spin populations injected from the contacts. A spin-up electron encounters a potential, which is proportional to the density of spin-down electrons, and vice versa. This leads to a repulsive interaction between electrons with opposite spin directions. Any external source leading to an imbalance between the density of spin-up and spin-down electrons is increased by the addition of the self-energy term $\Sigma_{int}^{\sigma}(x,y)$. In our case, it is the asymmetric LSOC, which leads to the initial imbalance. In the iterative scheme used to solve the Schrödinger equations for both spin populations and the Poisson equation, the asymmetry in the LSOC on both sides of the QPC eventually leads to a substantial increase in the imbalance between the spin-up and spin-down populations in the narrow portion of the QPC. This leads to a substantial difference in the self-energy term $\Sigma_{int}^{\sigma}(x, y)$ for both types of spins, with one becoming predominant for a specific range of common gate voltage ($V_G$) applied to the two SGs. Because $\Sigma_{int}^{\sigma}(x, y)$ is proportional to the strength $\gamma$ of the e-e interaction, the dominance of one-type of spin occurs only if $\gamma$ is strong enough. NEGF simulations illustrated this effect were described in details in refs. [10, 23, 31].

Unless otherwise specified, the strength of the parameter $\gamma$ in the interaction self-energy characterizing the strength of the electron-electron interaction was set equal to $3.7\hbar^2/2m^*$ and the parameter $\beta$ characterizing the strength of the LSOC was set equal 200 Å$^2$. For all simulations, $V_s = 0$V, $V_d = 0.3$mV. An asymmetry in the QPC potential confinement was introduced by taking $V_{sg1} = 0.2$ V $+ V_{sweep}$ and $V_{sg2} = -0.2$ V $+ V_{sweep}$ and the conductance of the constriction was studied as a function of the sweeping (or common mode) potential, $V_{sweep}$. At the interface between the rectangular region of size $w_2 \times l_2$ and vacuum, the conduction band discontinuities at the bottom and the top interface were modeled, respectively, as

$$\Delta E_c(y) = \frac{\Delta E_c}{2}\left[1+\cos\frac{\pi}{d}\left(y-\frac{w_1-w_2}{2}\right)\right], \qquad (2)$$

and

$$\Delta E_c(y) = \frac{\Delta E_c}{2}\left[1+\cos\frac{\pi}{d}\left(\frac{w_1+w_2}{2}-y\right)\right], \qquad (3)$$

to achieve a smooth conductance band change, where $d$ was selected to be in the nm range to represent a gradual variation of the conduction band profile from the inside of the quantum wire to the vacuum region. A similar grading was also used along the walls going from the wider part of the channel to the central constriction of the QPC (Fig. 6). This gradual change in $\Delta E_c(y)$ is responsible for the LSOC that triggers the spin polarization of the QPC in the presence of an asymmetry in $V_{sg1}$ and $V_{sg2}$. The parameter $d$ appearing in Eqns. (2) and (3) was set equal to 1.6 nm. The conductance (in units of $2e^2/h$) of the QPC was then calculated using the NEGF with a non-uniform grid configuration containing more grid points at the interface of the QPC with vacuum. All calculations were performed at a temperature $T = 4.2\ K$.

Figure 7 shows a plot of the conductance, $G$ of the QPC shown in Fig. 6 as a function of the common sweep voltage $V_{sweep}$ applied to the two SGs. The geometrical parameters ( in Fig. 6) used in the simulation domain are $w_2 = $ 16nm, and $w_1 = $ 48nm and $l_1 = l_2 + $ 32nm, with $l_2$ varying from 28 to 40 nm in steps of 2 nm. The different curves correspond to different values of $l_2$. A length $l_2$ equal to 28, 30, 32, 34, 36, 38, and 40 nm corresponds to an aspect ratio $l_2/w_2$ of the narrow portion of the QPC equal to 1.75, 1.875, 2.0, 2.125, 2.25, 2.375, and 2.5, respectively. We refer to $l_2/w_2$ as the geometrical aspect ratio since it is calculated as the ratio of the physical dimensions of the narrow portion of the QPC shown in Fig. 6. Another definition of the aspect ratio could be used in which a smaller value of $w_2$ would be used to take into account the effects of electrostatic confinement in the direction perpendicular to the direction of current flow ($y$ direction in Fig. 6). The electrostatically defined $w_2$ is smaller than the geometrical one which would lead to larger aspect ratio values. The corresponding values of $l_2/w_2$ would also increase as the length of the narrow portion of the QPC is increased in the direction of current flow.

Figure 7 shows that a conductance anomaly appears as a conductance peak below $2e^2/h$ followed by a negative differential region (NDR) as $l_2$ is increased from 30 to 32 nm. This NDR is a remnant of the conductance oscillations present in the conductance curve under the assumption of ballistic transport when the effects of electron-electron interaction are neglected and results from the multiple reflections at the ends of the narrow portion of the QPC. The NDR does not appear in the experimental results (shown in Fig. 4) because the fabricated QPCs have smooth edges, as shown in Fig. 1. This diminishes the effects of multiple reflections at the end of the narrow portion of the QPC. The first conductance peak appears at a lower value as the length of the narrow portion of the QPC increased, i.e., when the aspect ratio $l_2/w_2$ increases. The first conductance peak appears between $0.81\times(2e^2/h)$ and $0.39\times(2e^2/h)$ as $l_2$ is between 32 to 40 nm, which corresponds to an increase in aspect ratio from 2.0 and 2.5. The NEGF results are in qualitative agreement with the experimental results: the conductance anomaly around $0.5\times(2e^2/h)$ only appears in QPC with sufficiently large aspect ratio. The agreement is only qualitative because the simulated QPC must be kept much smaller than the actual device to reduce the time for the NEGF simulations to converge. The NEGF results suggest that the effects of electron-electron interactions (modeled as a repulsive Coulomb contact potential[23,24]) become prominent over an increased range of common sweep voltage as the QPC aspect ratio increases. The increase in aspect ratio changes the overall shape of the electrostatic confinement in the narrow portion of the QPC leading to an increase in the importance of the repulsive Coulomb contact potential in that region. This effect which becomes more prominent for QPCs with larger aspect ratio because the integrated spin densities are larger in these structures. This leads to different strength of the interaction self-energy, $\Sigma_{\text{int}}^{\sigma}(x,y)$, for spin-up and spin-down electrons, which ultimately favors the injection of one type of spin. As a result a spontaneous spin polarization occurs in the narrow portion of the QPC accompanied by a conductance anomaly near $0.5\times(2e^2/h)$[23,24].

To further investigate the importance of the strength of the electron-electron interaction on the formation, location, and overall shape of the conductance anomalies, the conductance of QPC device shown in Fig. 6 was computed for different values of the parameter $\gamma$ for a QPC device with the following dimensions: $l_2 = 40$nm, $l_1 = l_2 + 32$nm, $w_2 = 16$nm, and $w_1 = 48$nm. For this QPC, the first conductance peak appear around $0.35\times(2e^2/h)$ in Fig. 7. In the NEGF simulations, the following parameters were used: $V_{sg1} = 0.2$V + $V_{sweep}$ and $V_{sg2} = -0.2$V + $V_{sweep}$, $V_{ds} = 0.3$mV, T = 4.2K, , and $\beta = 200$ Å$^2$. The different curves correspond to values of $\gamma = 3.4$,

3.7, and 4.0 (in units of $\hbar^2/2m^*$), respectively. Figure 8 indicates that the parameter γ strongly affects the location of the first anomalous conductance peak and the size of the NDR following it.

To assess the importance of the onset of spin polarization in the narrow portion of the QPC, it is best to plot the conductance polarization α = [$G_\uparrow$ - $G_\downarrow$]/[$G_\uparrow$ + $G_\downarrow$] associated to the conductance curves as a function of the common sweep voltage $V_{sweep}$, where $G_\uparrow$ and $G_\downarrow$ are the conductance due to the majority and minority spin bands, respectively. Figure 9 is a plot α versus $V_{sweep}$ associated to the conductance curves shown in Fig. 8. Figure 9 clearly shows that a non-zero conductance polarization α appears over a larger range of common sweep voltage of the SGs as the strength γ of the electron-electron interaction increases.

The NEGF simulations described above show that the conductance anomalies only appear for a sufficiently large QPC aspect ratio. The agreement with the experimental results is only qualitative because the simulated QPC dimensions must be kept much smaller (by about a factor 10) than the actual device to reduce the time for the NEGF simulations to converge. The NEGF results suggest that the effects of electron-electron interactions (modeled as a repulsive Coulomb contact potential[23,24]) become prominent over an increased range of common sweep voltage as the QPC aspect ratio increases. It is the increase in the importance of the repulsive Coulomb contact potential with an increase of the QPC aspect ratio which eventually leads to the observation of a conductance anomaly near 0.5×(2$e^2$/h).

## 5. Conclusions

In the past, there have been several studies of the dependence of conductance anomalies (around 0.5×(2$e^2$/h) and 0.7×(2$e^2$/h)) on the length of the narrow portion of QPCs with *symmetrically biased* top gates to assess the importance of the effects of electron-electron interactions in these structures[16-19]. This work shows that the effects of electron-electron interactions, as mediated by an *asymmetry* in the LSOC of InAs QPCs with in-plane SGs, are responsible for a 4-fold increase in the range of common sweep voltage applied to the SGs over which the 0.5×($2e^2/h$) conductance plateau is observed when the QPC aspect ratio is increased by a factor 3. QPCs with in-plane SGs and with large aspect ratio can therefore achieve high spin injection efficiency spanned over a large range of common sweep voltage on the SGs. This is an essential step for the design of robust spin injectors and detectors which could be used for the realization of reliable spin-based field effect transistors. However, for this approach to become a reliable way to implement in more advanced spin-based circuits, the reliability and reproducibility of the results would have to first be tested on a large number of QPCs following the work described in refs. [20-22]

**Acknowledgment:** This work was partially supported by DST (SERB) Award EMR/2016/000808. P. P. D. is grateful to receive help and support from A. Das (RRCAT Indore, India). N. S. S. thanks NIT Karnataka and MHRD India for financial assistance.

Table 1: Details of the QPC geometrical dimensions, flatness $\Delta V_G^{Flat}$ of $0.5\times(2e^2/h)$ conductance plateau and relative increase of $\Delta V_G^{Flat}$ for different QPCs compared to its value for QPC 1.

| Device | Size of narrow portion of QPC (W, L) (in nm) | QPC aspect ratio (L/W) | $\Delta V_G^{Flat}$ (Volt) | Relative flatness of 0.5 plateau compared to QPC 1 |
|---|---|---|---|---|
| QPC 1 | 270, 320 | 1.2 | 0.3 | 1.0 |
| QPC 2 | 270, 380 | 1.4 | 0.5 | 1.7 |
| QPC 3 | 270, 510 | 1.9 | 0.8 | 2.7 |
| QPC 4 | 270, 820 | 3.0 | 1.0 | 3.3 |
| QPC 5 | 270, 930 | 3.4 | 1.2 | 4.0 |

## Figure Captions

**Figure 1:** Scanning electron micrograph (SEM) of a typical InAs QPC with two in-plane side gates (G1 and G2) created by cutting isolation trenches on a 2DEG using wet-etching technique. The current flows in the x-direction. An asymmetric bias is applied between the two side gates to generate a spin polarized current.

**Figure 2:** Schematic representation of the InAs/InGaAs quantum well structure used for fabricating QPC devices. The direction of epitaxial growth of constituent layers in the heterostructure is along z.

**Figure 3:** Circuit diagram for conductance measurements in a QPC channel defined by the two trenches shown as dark areas in figure. The drive voltage ($V_{DS}$), from voltage source $V_{osc}$, between the source (S) and drain (D) creates a current through the narrow channel. The current flows in x-direction. The conductance of the QPC channel is measured as a function of a common sweep voltage after a potential asymmetry is applied to the two SGs (G1 and G2). The voltage drop, $\Delta V$ across the channel (*i.e.*, between the two voltage probes (denoted by pads labelled $V_1$ and $V_2$) is measured with a digital voltmeter.

**Figure 4:** Conductance G (in units of *$2e^2/h$*) and transconductance ($dG/dV_G$) of five QPCs with different aspect ratios measured as a function of the common sweep voltage $V_G$ applied to the in-plane side gates. All measurements were performed at T = 4.2 K. The sweep voltage is superimposed on initial potentials $V_{G1}$ and $V_{G2}$ applied to the gates to create an asymmetric potential profile in the QPC constriction. The dimensions of QPCs 1 through 5 are given in Table 1.

**Figure 5:** Details of the extraction of the flatness of the conductance curve near the $0.5\times(2e^2/h)$ conductance plateau of QPC 5 through the use of the transconductance ($dG/dV_G$) as a function of common sweep voltage ($V_G$) applied to both side gates. The red horizontal line shows the minimum in the transconductance curve near the $0.5\times(2e^2/h)$ plateau. The green horizontal line is located at a factor 3 times the location of the red line leading to two points of intersection, $V_1$ and $V_2$, which are then extracted to calculate the flatness $\Delta V_G^{\text{Flat}} = |V_2 - V_1|$ of the 0.5 plateau.

**Figure 6:** Schematic of a QPC with in-plane SGs. The light gray areas represent the regions where the conduction band energy profile changes abruptly from the inside of the semiconducting channel to the vacuum region in the isolation trenches defined by wet chemical etching (dark areas). In the narrow portion of the QPC (of width $w_2$ and length $l_2$) the sharp potential discontinuity on the sidewalls leads to LSOC. A bias asymmetry between the SGs eventually leads to spin polarization in the narrow channel of the QPCs. Also shown are the source and drain contacts. The current flows in the x-direction.

**Figure 7:** Conductance (in units of $2e^2/h$) $G$ of asymmetrically biased QPC as a function of the common sweep voltage $V_{\text{sweep}}$ applied to the two SGs of the QPC shown in Fig.6. The potential on the two SGs are $V_{\text{sg1}} = 0.2\text{V} + V_{\text{sweep}}$ and $V_{\text{sg2}} = -0.2\text{V} + V_{\text{sweep}}$. The curves labeled 1 through 7 correspond to $l_2 = $ 28, 30, 32, 34, 36, 38, and 40 nm, respectively. The physical dimensions of the simulation domain were set equal to $l_1 = l_2 + 32\text{nm}$, $w_2 = 16\text{nm}$, and $w_1 = 48\text{nm}$. In all NEGF simulations, the following parameters were used: $V_{\text{ds}} = 0.3\text{mV}$, T = 4.2K, γ = 3.7 in units of $\hbar^2/2m^*$, and β = 200 Å$^2$.

**Figure 8:** Conductance (in units of $2e^2/h$) $G$ of asymmetrically biased QPC as a function of the common sweep voltage $V_{\text{sweep}}$ applied to the two SG of the QPC shown in Fig.6. The potential

on the two SGs are $V_{sg1} = 0.2V + V_{sweep}$ and $V_{sg2} = -0.2V + V_{sweep}$. Furthermore, $V_{ds} = 0.3mV$, T = 4.2K, and β = 200 Å². The different curves correspond to values of γ = 3.4, 3.7, and 4.0 (in units of $\hbar^2/2m^*$), respectively.

**Figure 9:** Plot of the conductance polarization α = $[G_\uparrow - G_\downarrow]/[G_\uparrow + G_\downarrow]$ as a function of the common sweep voltage $V_{sweep}$ applied to the two SGs of the QPC shown in Fig.6. The potential on the two SGs are $V_{sg1} = 0.2V + V_{sweep}$ and $V_{sg2} = -0.2V + V_{sweep}$. The physical dimensions of the simulation domain were set equal to $l_2$ = 40nm, $l_1 = l_2$ + 32nm, $w_2$ = 16nm, and $w_1$ = 48nm. The following parameters were used: $V_{ds}$ = 0.3mV, T = 4.2K, and β = 200 Å². The three different curves correspond to values of γ = 3.4, 3.7, and 4.0 in units of $\hbar^2/2m^*$.

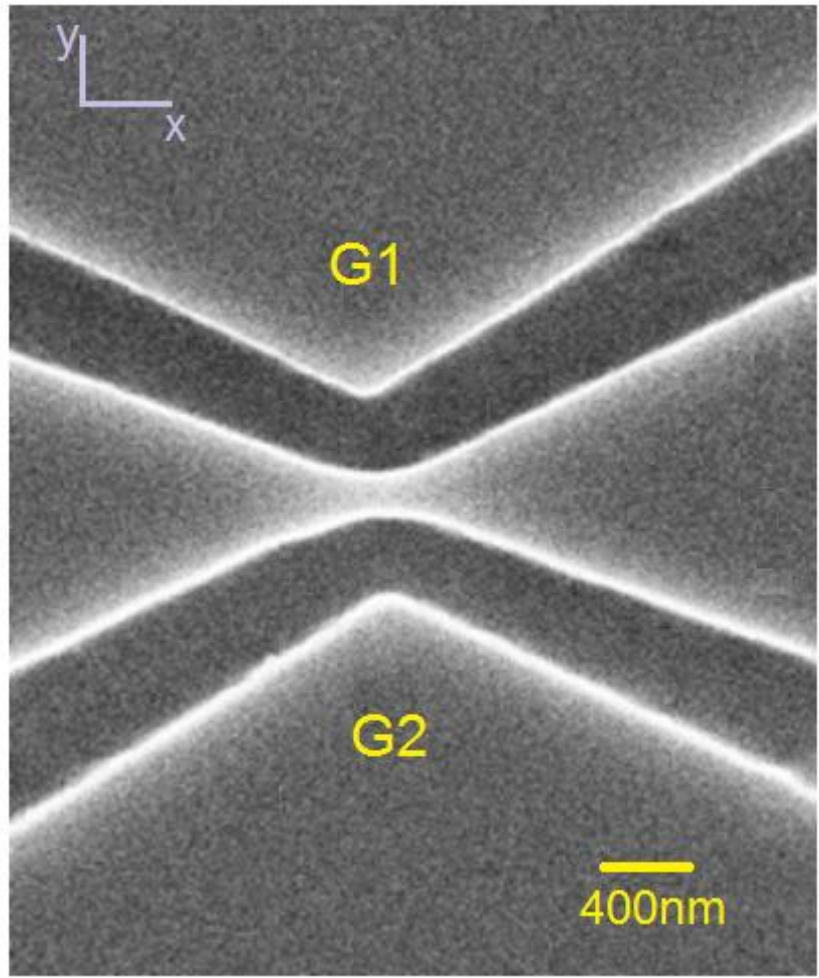

**Fig 1 (P. Das et al)**

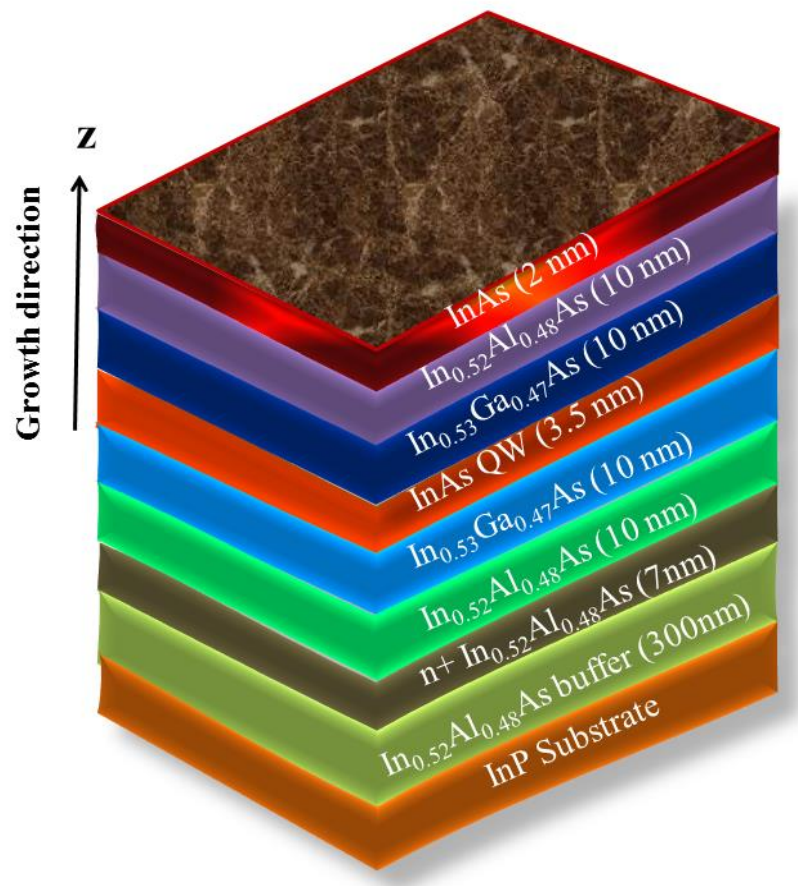

**Fig 2 (P. Das et al)**

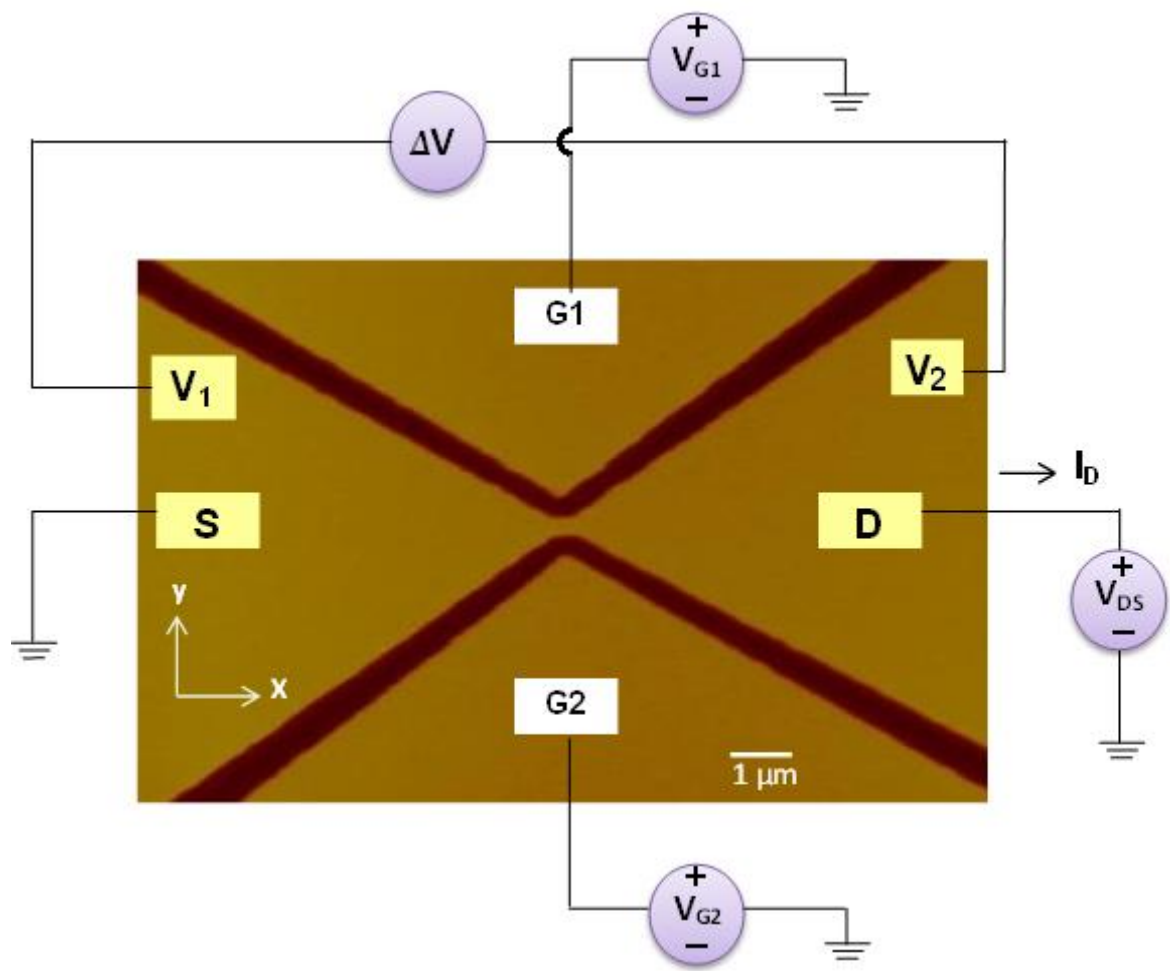

**Fig 3 (P. Das et al)**

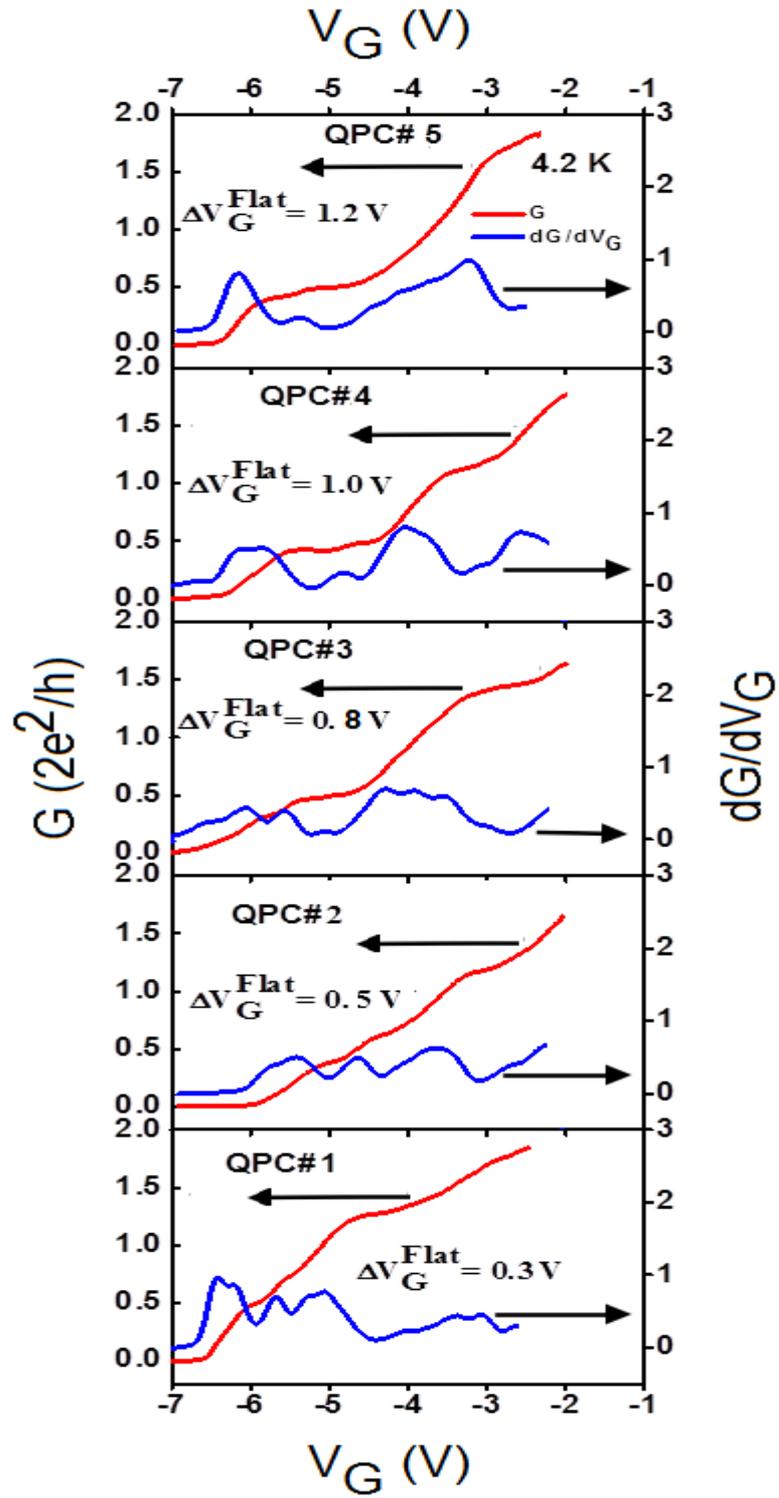

**Fig 4 (P. Das et al)**

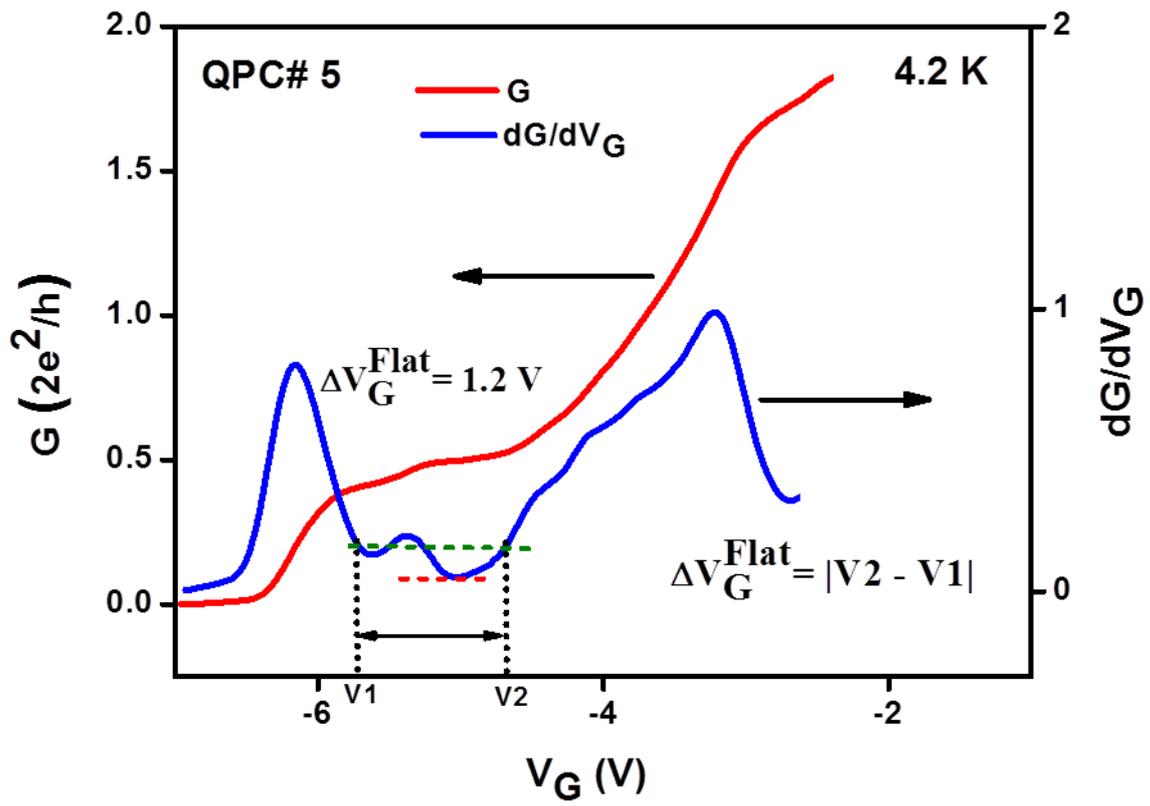

Fig.5 (P. Das et al)

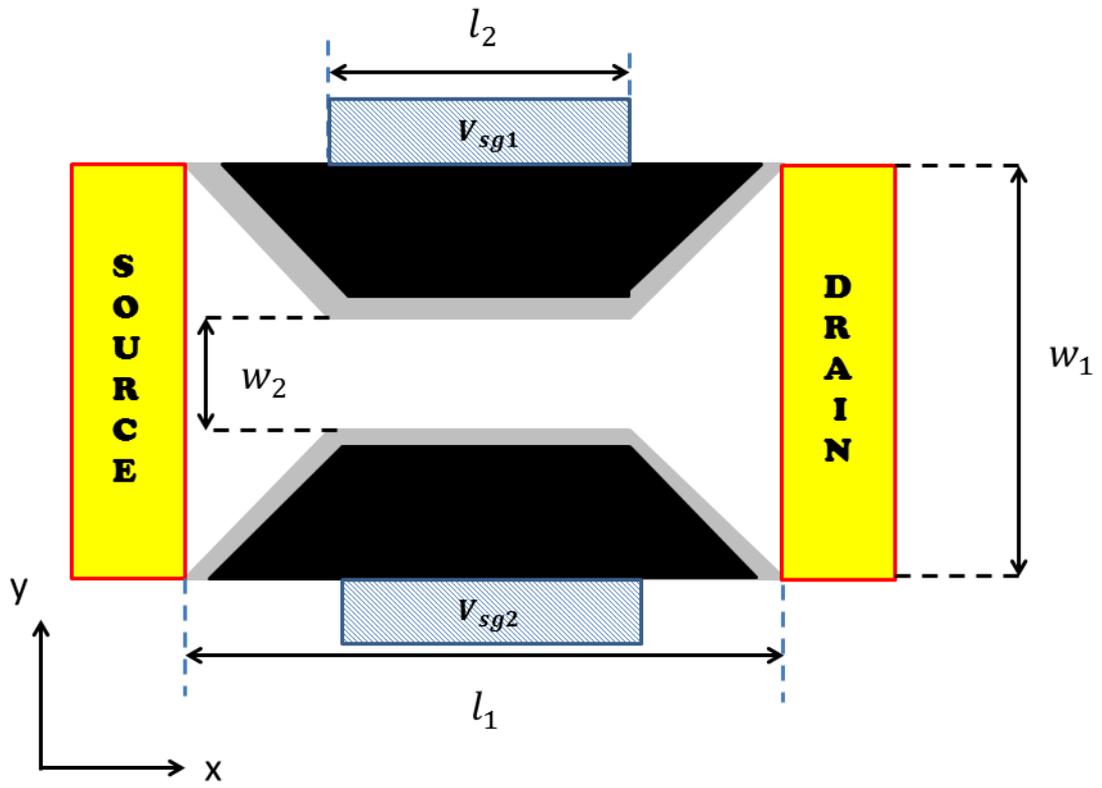

**Fig. 6 (P.Das et al)**

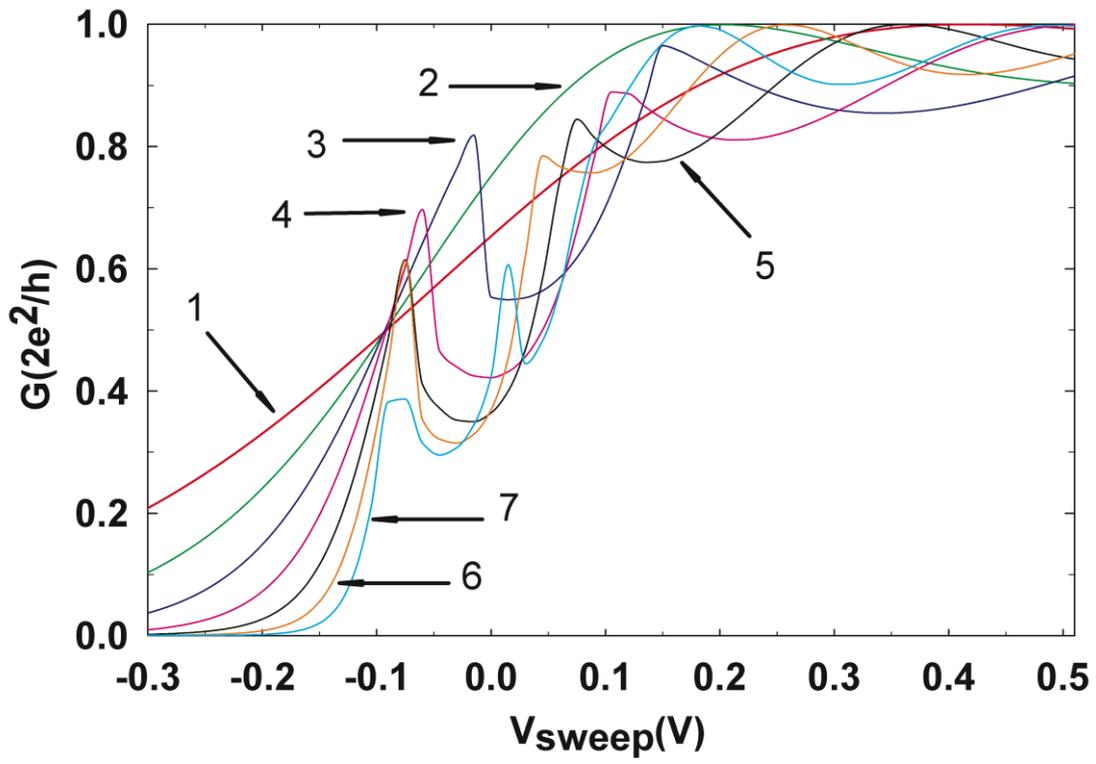

**Fig.7 (P. Das et al)**

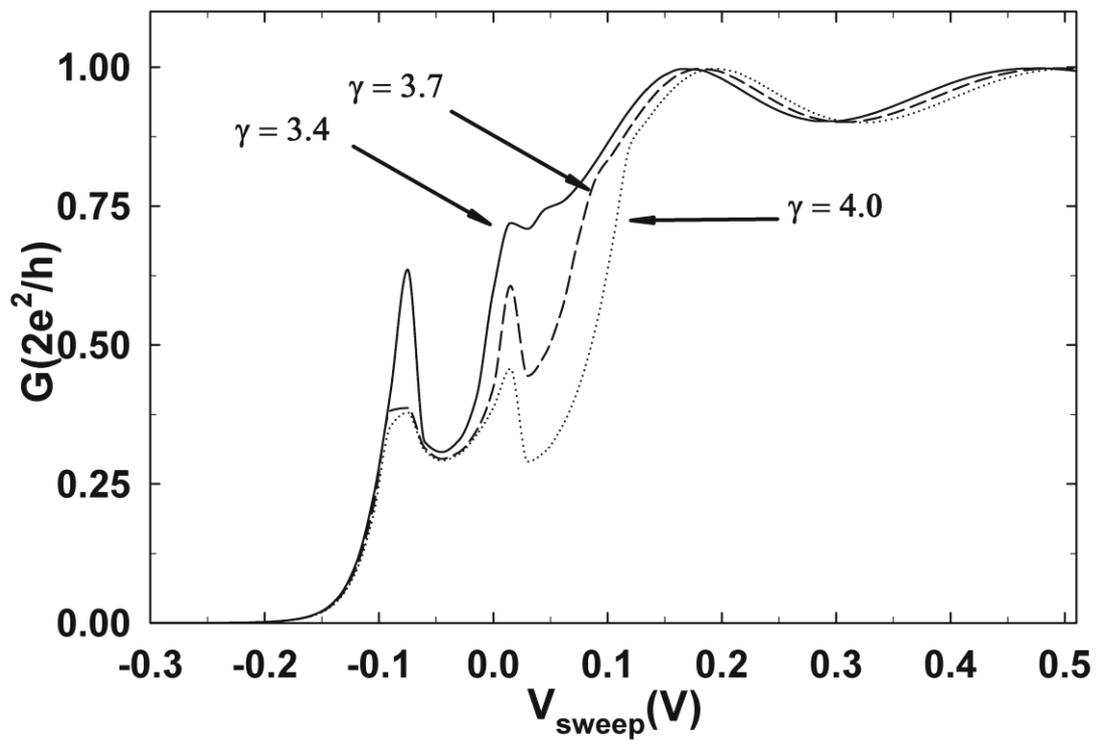

Fig.8 (P. Das et al)

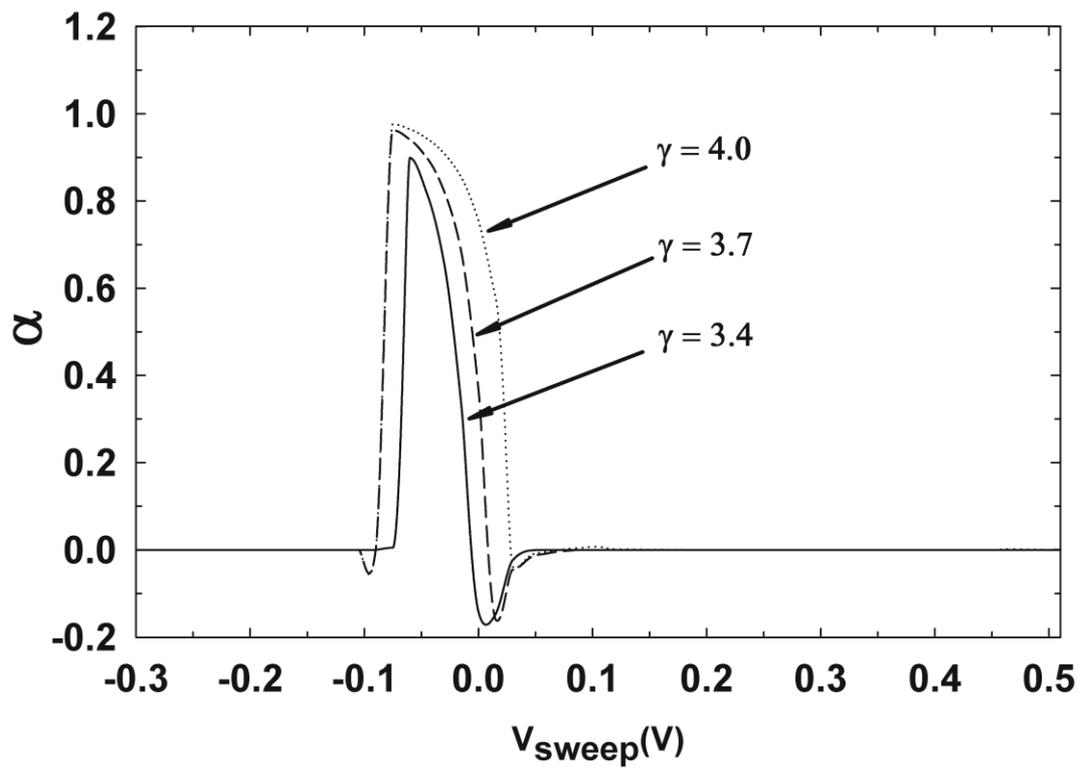

Fig.9 (P. Das et al)